 \documentclass[preprint,review,12pt]{elsarticle}

\usepackage{amsmath}
\usepackage{amssymb}
\usepackage{graphicx}
\usepackage{subcaption}
\usepackage{multirow}
\usepackage{framed}
\usepackage{afterpage}
\usepackage{verbatim}
\usepackage{psfrag}
\usepackage{setspace}
\usepackage{epsfig}
\usepackage{epstopdf}
\usepackage{footmisc}
\usepackage{fmtcount}
\usepackage{stackrel}
\usepackage{float}
\usepackage{breqn}
\usepackage{enumerate}
\usepackage{longtable}
\usepackage{hyperref}
\usepackage{hypcap}
\hypersetup{pdftex,colorlinks=true,allcolors=blue}
\usepackage{colortbl}
\usepackage[utf8x]{inputenc} 
\usepackage{color}
\journal{Physical Communication}
\begin{document}

\begin{frontmatter}

\title{ Optical Camera Communications: Survey, Use Cases, Challenges, and Future Trends
}
\tnotetext[label1]{Corresponding Author: Nasir Saeed (mr.nasir.saeed@ieee.org)}
\author{Nasir Saeed, Shuaishuai Guo, Ki-Hong Park, Tareq Y. Al-Naffouri, Mohamed-Slim Alouini}
\address{Computer, Electrical and Mathematical Sciences \& Engineering (CEMSE) Division, King Abdullah University of Science \&  Technology (KAUST), Thuwal, KSA.}

\begin{abstract}
Recently, a camera or an image sensor receiver based optical wireless communications  (OWC) techniques have attracted particular interest in areas such as the internet of things, indoor localization, motion capture, and intelligent transportation systems.  As a supplementary technique of high-speed OWC  based on photo-detectors, communications hinging on image sensors as receivers do not need much modification to the current infrastructure, such that the implementation complexity and cost are quite low. Therefore, 
in this paper, we present a comprehensive survey of optical camera communication (OCC) techniques, and their use in localization, navigation, and motion capture. This survey is distinguishable from the existing reviews on this topic by covering multiple aspects of OCC  and its various applications. The first part of the paper focuses on the standardization, channel characterization, modulation, coding, synchronization, and signal processing techniques for OCC systems while the second part of the article presents the literature on OCC based localization, navigation, motion capture, and intelligent transportation systems.
Finally, in the last part of the paper, we present the challenges and future research directions of OCC.

\end{abstract}

\end{frontmatter}

\section{Introduction}
The demand for mobile data communications is overgrowing with the pervasive connectivity of the Internet of Things (IoT) and the growth of digital services such as social media and video contents.  According to the latest forecast report of Ericsson, it is expected that the total mobile traffic will rise with annual growth rate of 42\% \cite{Ericsson2017}. In 2023, monthly global mobile data traffic will surpass 100 exaBytes (EB) \cite{Ericsson2017}.  
To satisfy this exponentially increasing demand, either increasing the bandwidth or improving the spectral efficiency should be utilized. However, the increase in spectral efficiency is slow and is unable to meet this insatiable demand. Exploiting new spectrum becomes a unique solution, and a much wider spectrum bandwidth of terahertz (THz) is in demand. 
Under such circumstances,  optical wireless communications (OWC)  have attracted great research interest in recent years due to its many desirable properties, including a large amount of available spectrum (from 350 nm to 1550 nm), high energy-efficiency, independent regulation, and well-controlled communication security \cite{Wu2014,Pathak2015,Ghassemlooy2016}.

Hence, OWC appeared as an alternative and a complementary option to the existing radio frequency (RF) communications. 
Based on the utilization of the spectrum, current research on terrestrial OWC can be divided into four categories: free space optical (FSO) communications, visible light communications (VLC),  light-fidelity (Li-Fi), and optical camera communications (OCC) \cite{Chowdhury2018}. FSO communication systems consist of a laser diode (LD) transmitter and a photodiode (PD) receiver. It typically relies on UV or visible bands, offers high rate transmission in a long distance, and can be used for a backhaul of communication networks.  However, FSO communications need a strict alignment, and consequently, their cost at transceivers is high.  Also, FSO communication systems suffer from atmospheric turbulence which can be mitigated by using different statistical channel models \cite{Sharma2014, Sharma2015, Sharma2017, Joshi2016, Sharma2018}, robust modulation techniques \cite{Amhoud2019, Trichili2019}, and accurate pointing and tracking methods \cite{Bashir2016, Bashir2017, Kaymak2018}. On the other hand, VLC hinging on visible bands is an LD or light emitting diode (LED) transmitter and PD receiver-based medium-range communication technology. VLC is capable of offering high data rate within a range of tens of meters but does not consider multiple user access. The advancement of VLC systems has also led to enable various location aware indoor applications.  Recently, several indoor localization systems based on VLC are proposed  \cite{Kuo2014, Yang2015, Li2014, Xie2016, Sheoran2018}. Interested readers are referred to \cite{Luo2017, Saeedsurvey2019}, and the references therein for VLC based indoor localization and tracking methods. Nevertheless, the VLC technology suffers from both limited coverage and interference. Increasing the field of view of the LEDs improves the coverage; however, it increases the interference at the receiver. Signal-to-interference ratio based methods can be used to analyze the coverage and interference problems in VLC systems \cite{Sharma2019}.

Different from VLC,  Li-Fi is an LED transmitter based light networking technology that involves multiple user access, bidirectional communications, multi-cell handover, etc. The emerging Li-Fi supports mobile communications, enables seamless access and can offer a speed of megabits per second (Mbps). However, all the receivers of FSO, VLC, and Li-Fi technologies consist of PDs, which rarely exists in current receiver devices and the commercialization cost to change the receiver structure is high. Therefore, developing a practical OWC system to harness its immediate benefits using commercial off-the-shelf devices is still an open issue \cite{Hao2016a}. This motivates the concept of OCC.  Different characteristics and limitations of OCC and VLC are presented in Table \ref{Tablevlcocc}.

\begin{figure}
  \centering
  \includegraphics[width=1\textwidth,height=0.45\textheight]{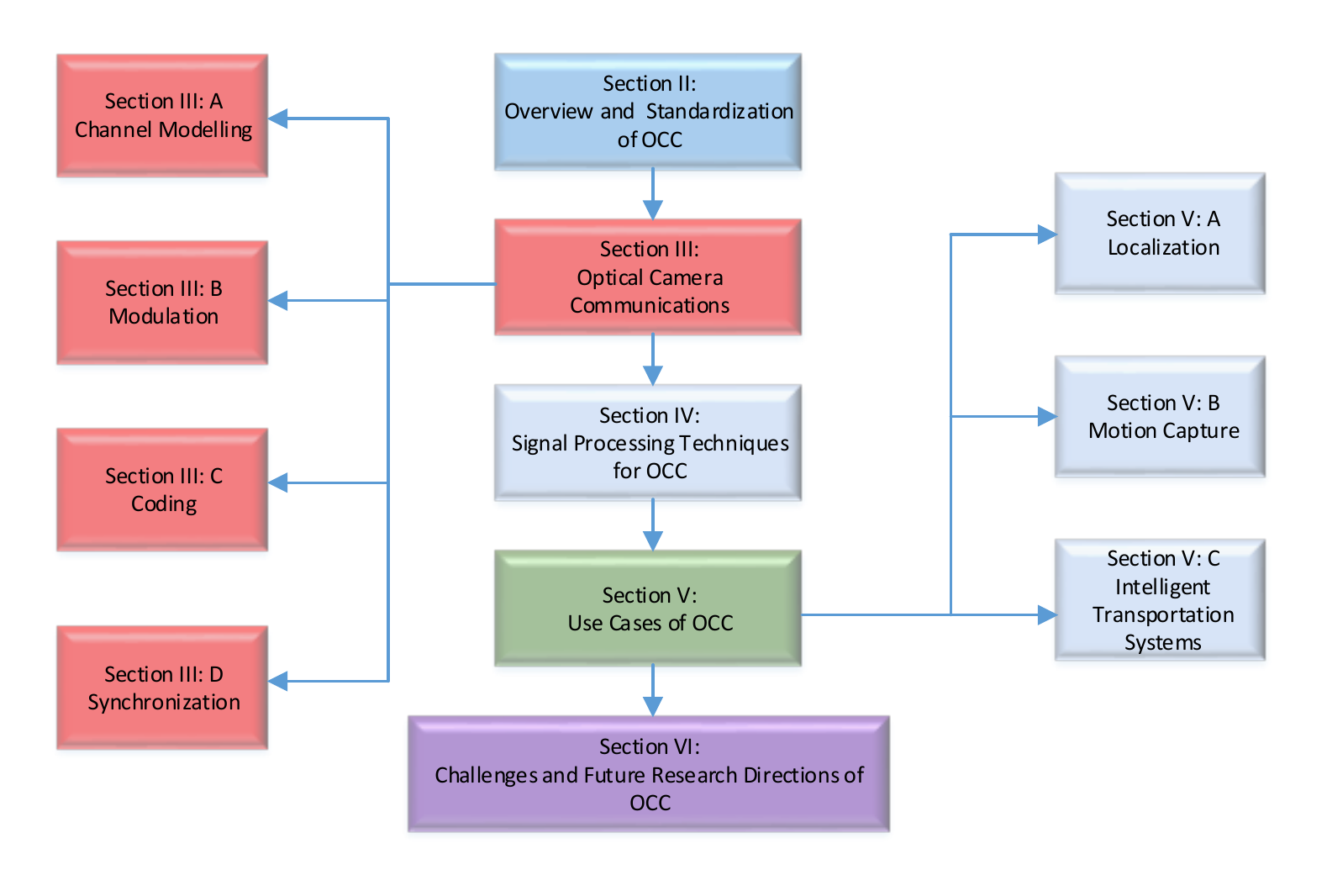}\\
  \caption{Organization of the main body of the survey paper.}
  \label{OCC_Organization}
\end{figure}
OCC is a communication technology that utilizes optical image sensors as receivers based on IR or visible bands, and thus it is also referred to as image sensor communications \cite{Boubezari2016, Nguyen2018a}.
Compared to other types of OWC technologies, OCC has many advantages.  First, it does not need any modification on the receiver where current off-the-the-shelf smartphones, digital cameras,  rear vehicle cameras, and surveillance cameras are all possible receiver candidates. Thus, OCC can facilitate a large body of applications. Second, the receivers of OCC are based on millions of pixels, and these pixels provide large degrees of freedom (DoFs) to transmit data and to handle the access of a massive number of users. Third, the image sensors of current cameras can usually deal with three colors which enable the transmission over the color domain. Owing to these advantages in cost, popularity, and information carrying capability, OCC has attracted much interest in areas such as IoT, indoor localization, motion capture, and intelligent transportation systems (ITS).  However, OCC has its limitations which include but not limited to low data rate due to the low sampling rate at the receiver, out-of-focus effect, unstable frame rate, and random block. The research community on OCC has proposed different solutions to overcome the above limitations. Therefore, the central theme of this survey is to introduce the novice readers about various limitations of OCC and their solutions. 
Furthermore, we provide technical details of OCC such as channel modeling, modulation, coding, and synchronization. Also, we survey the literature on localization and motion capture systems developed based on OCC. Besides, challenges and future research directions for OCC are provided.

\begin{table}
\footnotesize
\centering
\caption{Comparison of OCC and VLC.}
\label{Tablevlcocc}
\begin{tabular}{|p{2.8cm}|p{2.8cm}|p{2.4cm}|}
\hline
\hline
\textbf{Characteristics}              & \textbf{OCC}         & \textbf{VLC}  \\ \hline \hline
Transmission range                     & Up to several km  &  Low \\ \hline 
Wavelength                       & UV, IR, and Visible light          & Visible light \\ \hline 
SNR                      & High         & Low      \\ \hline 
Receiver                      & Camera           & Photo-detector \\ \hline 
Decoding                      & High complexity   & Low complexity \\ \hline 
Data rate                      & Lower than VLC (in kbps)       & 11.67 kbps - 96 Mbps      \\ \hline 
Protocol                     & IEEE.802.15.7r1       &  IEEE.802.15.7        \\ \hline
MIMO Implementation                      & Easy       &  Difficult        \\
\hline
\hline        
\end{tabular}
\end{table} 

\subsection{Related Surveys} 
The increasing demands of OCC applications lead to the publications of a few brief surveys on modulation and coding techniques for OCC. Most of the existing surveys on OCC have been focusing on modulation techniques. For example, in \cite{Bae2017}, the history, modulation technique, advantages, and limitations of OCC are briefly introduced. Furthermore, the paper is focused on addressing the weaknesses of OCC such as camera sampling rates, frame rate variations, and motion stabilization. Similarly,  in \cite{Saha2015}, Saha \emph{et al}  presented a survey focusing on the vital technology consideration in IEEE 802.15.7r1 task group for OCC systems. In \cite{Islam2017}, the authors have presented a survey on the applications of OCC in intelligent transportation systems (ITS), including vehicular to vehicular (V2V) communications, vehicular to infrastructure (V2I) communications,  cloud-based internet of vehicular communications (IoV), etc. A comparative study of different OWC systems has been presented in \cite{Chowdhury2018} which also include possible architectures and applications of OCC. The latest survey works dedicated to OCC was presented in \cite{Nguyen2018,Le2017, Luo2018}. In \cite{Nguyen2018}, the authors reviewed OCC proposals of IEEE 802.15.7m task group, showed the essential technical consideration of the OCC specifications, and discussed the future research directions. A similar approach was followed in \cite{Le2017} where the authors have presented the design and implementations issues of OCC. Furthermore, the authors in \cite{Luo2018}, reviewed different modulation techniques and discussed the specialized frame structures for each modulation technique.
 
Comparison between all of the surveys mentioned above is presented in Table \ref{Tablerelatedsurveys}. In this survey, we cover the existing literature on OCC, including the standardization, channel characterization, modulation, coding, synchronization, localization, navigation, motion capture, and intelligent transportation systems. We believe that including all these aspects of OCC into a survey form makes our work a unique contribution to the research communities working on OCC, localization and motion capture using OCC, and OCC-based intelligent transportation systems.

\begin{table}
\footnotesize
\centering
\caption{Comparison of this paper with the existing surveys.}
\label{Tablerelatedsurveys}
\begin{tabular}{|p{2.1cm}|p{0.9cm}|p{6.9cm}|}
\hline
\hline
 \textbf{Ref.}              & \textbf{Year}         & \textbf{Area of Focus}  \\ \hline\hline 
Saha \emph{et al} \cite{Saha2015}                     & 2015  & Design and implementation  \\ \hline 
Bae \textit{et al.} \cite{Bae2017}                      & 2017          & Receiver architectures \\ \hline 
Islam \textit{et al.} \cite{Islam2017}                      & 2017           & OCC for ITS      \\ \hline 
Le \textit{et al.} \cite{Le2017}                      & 2017           & Design and implementation          \\ \hline 
Chowdhury \textit{et al.} \cite{Chowdhury2018}                      & 2018   & Architectures and applications \\ \hline 
Nguyen \textit{et al.} \cite{Nguyen2018}                      & 2018       & Modulation and coding      \\ \hline 
Luo \textit{et al.} \cite{Luo2018}                      & 2018       &  Modulation and coding        \\ \hline
This paper                      & 2019       &  Standardization, channel characterization, modulation, coding, synchronization, localization, navigation, motion capture, and future research directions        \\
\hline
\hline        
\end{tabular}
\end{table}

\subsection{Survey Organization}
The survey is organized as illustrated in Fig. 1. Section II presents the overview, history, and standardization of OCC. Section III provides a comprehensive taxonomy on OCC technologies including channel modeling, modulation and coding techniques, and system synchronization methods. Section IV and V presents the literature on signal processing techniques and use cases of OCC, respectively. In Section VI, we articulate the challenges and future research directions related to OCC. Finally, Section VIII concludes the paper.   
\begin{figure*}[t]
  \centering
  \includegraphics[width=1\textwidth,height=0.2\textheight]{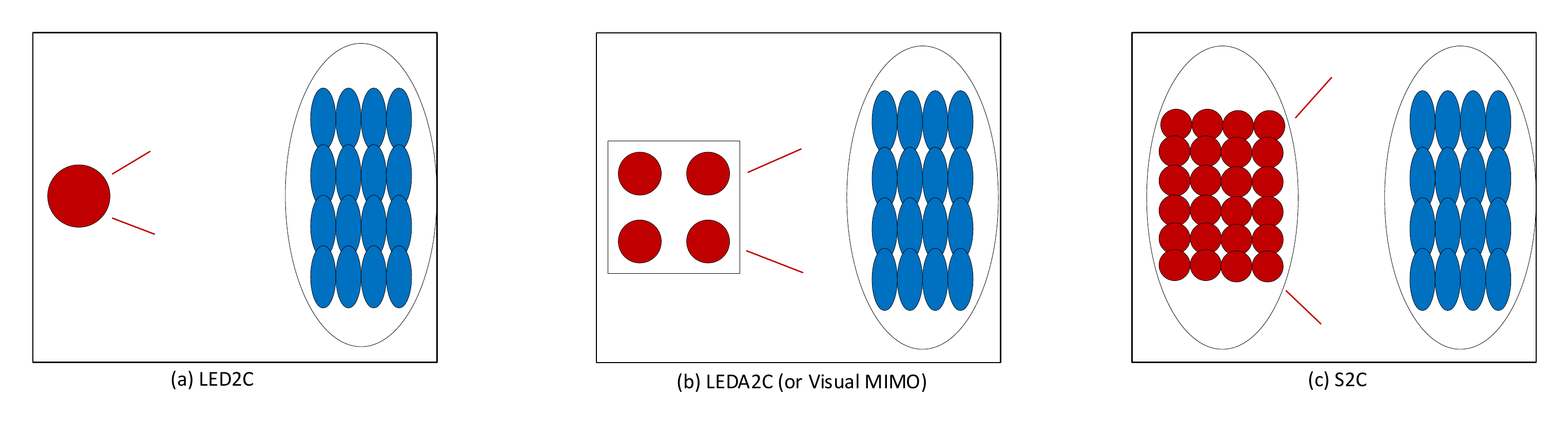}\\
  \caption{Three categories of OCC.}
  \label{Category}
\end{figure*}
\section{Brief Overview and Standardization of OCC}
In this section we provide an overview on the development of OCC system in chronological order. Furthermore, various standardization activities for OCC are presented. 
\subsection{Overview}

Recently, OCC has emerged as a new technology for the existing VLC systems. The comparison of  OCC and VLC is presented in Table \ref{Tablevlcocc} where both the systems have their pros and cons. However, the advantages of low cost and large transmission range make OCC  unique in the  OWC family. The history of OCC dates back to 2001 when Leibowitz \emph{et al} firstly presented FSO communications with the camera as a receiver \cite{Leibowitz2001}. Since the past decade, a lot of research has been done on OCC to improve the data rate, increase the transmission range, and to develop the efficient design and implementation schemes. The standardization group IEEE 802.15.7 was found in 2011 to provide a specification for the physical (PHY) and medium access control (MAC) layers for VLC which lead to the standardization of IEEE 802.15.7m  for  OCC.
\begin{figure*}[t]
  \centering
  \includegraphics[width=1\textwidth,height=0.42\textheight]{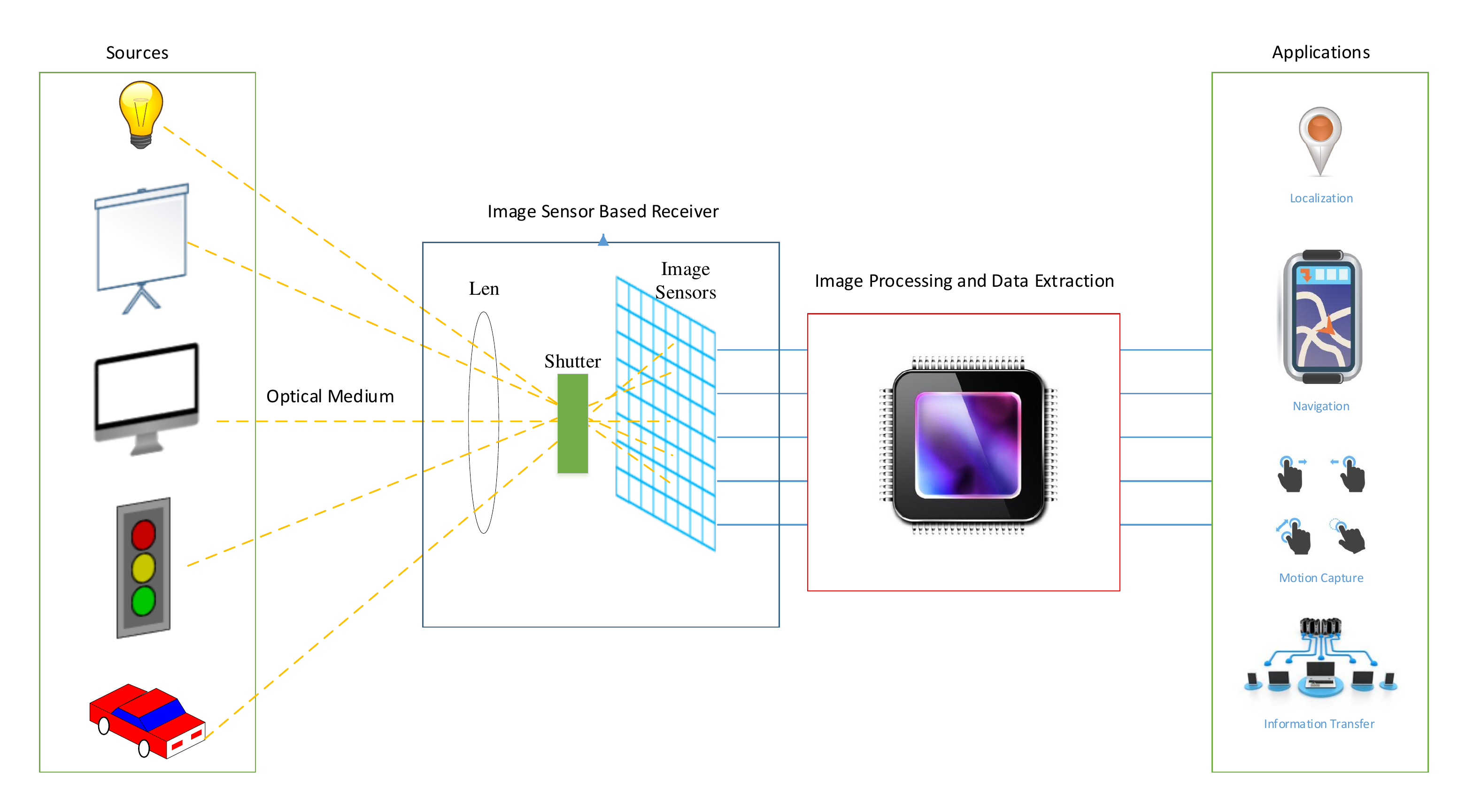}\\
  \caption{Optical camera receiver structure, communication procedure and potential applications.}
  \label{OCC_Receiver}
\end{figure*}

As a standalone technique, OCC can be divided into three categories as illustrated in Fig.~\ref{Category}. The first category is single LED as a transmitter and camera as a receiver (LED2C),  the second category consists of the LED array as a transmitter and camera as a receiver  (LEDA2C) which is also referred to as visual MIMO \cite{Ashok2010}. The third category consists of a screen as a transmitter and camera as a receiver (S2C) \cite{Wengrowski2016}. 

Based on the above classification, the possible candidates for a transmitter in OCC systems are LEDs, Television screens, projector screens, monitor screens, traffic lights, and the car lights (Fig. \ref{OCC_Receiver}). The light from all these different sources passes through the wireless channel and arrives at the lens of the receiver. Afterward, the light passes through the shutter, which controls the exposure of the image sensor/camera. There exist two different types of the shutter, the global and rolling shutters. The global shutter controls the image sensors simultaneously, while the rolling shutter controls the image sensors to expose row by row \cite{Le2017}. An image of the source captured is generated at the image plane of the camera, which can further be processed using image processing techniques to extract the data required for various applications. At the image sensors, the light is first transformed to the electronic signal, then quantized to an image,  and finally compressed to a specific image format. By image processing and data extraction schemes, the information carried by space/time/color domain of the light can be decoded for localization and navigation. The variation of the light channel can be used for motion recognition. Even though the data rate is low due to the limited frame rate, it can convey some location-aware information at a low speed, and can also perform secure near field communication.

\subsection{Standardization}

The IEEE standards which cover the optical spectrum are IEEE 802.15.7-2011 (VLC), IEEE.802.15.7m (OCC), and IEEE 802.15.13 (Li-Fi). The standard of IEEE 802.15.7-2011 is well established for VLC systems.  The release of a standard for VLC leads the researchers to investigate OCC which also complements the existing RF technologies and does not interfere with it. Therefore, it was required to develop a task group to come up with the standardization of OCC which includes the OCC transceivers, system architecture, PHY and MAC layers specification. A task group called TG7m was founded in 2014 to develop the technical requirements of OCC, and it is expected that the standardization of IEEE.802.15.7m will be completed by the end of 2018. Latest updates on the standardization of OCC are available in \cite{IEEEmentor2018}. Currently, there are eight different standards for PHY layer modes proposed for OCC while the MAC layer is still under discussion. Each PHY layer mode is defined based on a different modulation scheme. The MAC layer of OCC is supposed to support the MAC layer modes of VLC, i.e., unidirectional data transfer, identification (ID) broadcast, and bidirectional data transfer. For example, the authors in \cite{Le2016b} suggested the use ID broadcasting mode for OCC. The interested readers are referred to \cite{Nguyen2018a} for standardization of OCC and achievements of TG7m. 

By using the existing or minimally modified hardware,  OCC will facilitate the development of interoperable and economically attractive products/solutions for a broad range of applications such as indoor positioning and navigation, augmented reality, and IoT applications.

\section{Optical Camera Communications}
Optical camera communications have recently got much attention due to the phenomenal technology advancements in smart devices. To date, several prototypes based on OCC are developed. However, it has not been deployed commercially yet. In this section, we cover physical layer aspects of OCC which include channel modeling, modulation, and coding.
\subsection{Channel Modeling}

Generally, intensity modulated direct detection (IM/DD) scheme is used for OCC where the information signal modulates the light intensity for the transmission. The information signal is non-negative and is proportional to the intensity of the light. On the receiver side, a pixel is used as a basic unit for the power detection which represents the photon count rate at the receiver area. The received signal is affected by multiple phenomena such as scattering and different types of noise. The photons arrival rate at the receiver is a random process which is modeled by the Poisson distribution \cite{optbook}. The intensity of the received signal is given as
\begin{equation}
I_y = h X + N_b,
\end{equation}
where $h$ is the channel gain, $X$ is the transmitted signal, and $N_b$ is the background noise intensity. In ideal case $h =1$ and $N_b = N_s+ N_{p}+N_r$, where $N_s$ is the shot noise for photons and dark current, $N_p$ is the noise due to non-uniform photo-response in pixel output, and $N_r$ is the read noise which is a collection of noise independent of the signal. Then, the received signal in electrical form can be written as
\begin{equation}
Y = X + Z_b,
\end{equation} 
where $Z_b$ represents the background noise in electrical form. The signal to noise ratio (SNR) of the received signal is written as
\begin{equation}\label{eq: snr}
\gamma = \frac{\textit{E}\{X^2\}}{\sigma_b^2},
\end{equation}
where $\textit{E}\{X^2\} = (I_p t)^2$ and $\sigma_b^2 = \sigma_s^2 +\sigma_p^2+\sigma_r^2$ is the variance of the background noise. $t$ is the integration time in which $I_p$ number of photons are aggregated by each pixel. In the following, we describe the variances for each type of noise. The variance of shot noise is given by
\begin{equation}
 \sigma_s^2 = q t (I_p + I_b + I_d),
\end{equation}
where $q$ is the quantization step, $I_b$ is the induced current due to background radiation, and $I_d$ is the dark current. Similarly, the variance of noise due to non-uniform photo-response is given as
\begin{equation}
 \sigma_p^2 = \kappa\bigg((I_p t)^2 + (I_b t)^2\bigg).
\end{equation}
where $\kappa$ is the factor value for photo-response non-uniformity. The read noise $N_r$ consists of an analog to digital conversion noise (AN), reset noise (RN), and source follower noise (SN), and therefore has the following variance
\begin{equation}
\sigma_r^2 = q^2 (\sigma_{AN}^2 + \sigma_{RN}^2 + \sigma_{SN}^2).
\end{equation}
Putting the values of $\sigma_s^2 $, $\sigma_p^2$, and $\sigma_r^2$ in \eqref{eq: snr} yields the SNR for OCC which depends on the pixel area, photo-current, integration time, and different type of noise sources. Moreover, the SNR can be improved by leveraging the spatial diversity (mapping of a single transmitter by combining multiple pixels into a block). In \cite{Ashok2010}, the authors have taken the advantage of spatial diversity by using visual MIMO to improve the SNR, given as 
\begin{equation}\label{SNR}
\gamma= \begin{cases}\frac{k P_t^2 d^{-2}}{e R P_n W f^2 l^2},~~~\textrm{if}~ d< d_c \\\frac{kP_t^2d^{-4}}{e R P_n A W s^2},~~~~\textrm{if}~ d\geq d_c,
\end{cases}
\end{equation}
where $k$ represents the LED's parameters such as irradiance angle, optical gain, the field of view, and Lambertian radiation pattern. $P_t$ represents the transmission power of a single LED, $d$ is the distance between the LED and the camera, $e$ is the electron charge density, $R$ is  the camera  responsitivity, $P_n$ is the power of background light, $W$ is the sampling rate of camera, $f$ is the focal length of the camera, $l$ is the diameter of LED, $s$ is the edge length of a pixel, and $d_c = \frac{fl}{s}$ is the critical distance at which the LED generates an image which falls onto a single pixel. It is clear from \eqref{SNR} that when $d < d_c$ there is a  gain of the order of $d^2$ in the SNR of the camera receiver as compared to a single photo-diode receiver. In case of $d \geq d_c$ there is no such gain and the SNR is equal to the case of a single photo-diode receiver.

Based on the SNR calculation in \eqref{SNR}, the capacity of OCC is given by Shannon formula in \cite{Ashok2010} as
\begin{equation}
C=W\bigg(W_s\log_2(1+\gamma)\bigg),
\end{equation}
where $W_s$ is the spatial bandwidth, which is equivalent to the number of orthogonal or parallel channels. Based on the above formula, in \cite{Ashok2010} Ashok \emph{et al}  analyzed the exact channel capacity of S2C communications where the results indicated that there is a room for at least 2.5 times improvement in the throughput of the existing S2C prototypes.

\subsection{Modulation Techniques}
\begin{figure}
  \centering
  \includegraphics[scale=0.7]{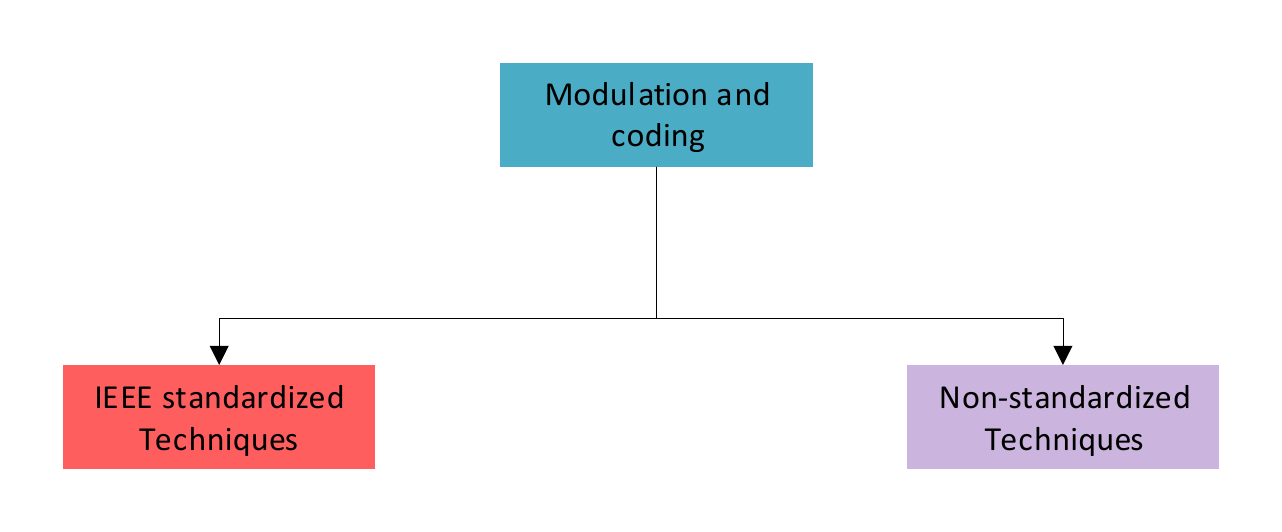}\\
  \caption{Classification of modulation and coding techniques for OCC.}\label{fig:modulation}
\end{figure}

There are numerous modulation techniques developed for OCC, to name a few, twinkled variable pulse position modulation (VPPM),  spatial 2-phase shift keying (S2-PSK), and  pulse width modulation (PWM).  However, the main obstacle for all of the OCC based modulation techniques is flickering. The existing flicker-free modulation techniques are limited by serving a camera at a short distance. 
The modulation techniques for OCC can be classified into the following two categories (see Fig.~\ref{fig:modulation}) based on the IEEE standardization:
\subsubsection{IEEE Standardized Techniques}
One of the most widely used modulation techniques for OCC is under-sampled frequency shift on-off keying which was first proposed by Intel \cite{Roberts2013,Roberts2013a,Roberts2013b}. It employs two different high frequencies to simultaneously modulate data and avoid flicking at the transmitter while the receivers do the sampling in a low frame rate. Similarly, Intel also developed twinkled VPPM for OCC where the modulated signals consist of two signaling sources, i.e., a high and a low frame rate source. The low frame rate camera is used to locate the twinkle LEDs, which modulate the data while the high frame rate camera is adopted for variable pulse position modulated data. The duty cycle is variable in the low frame rate to generate the fickle amplitude and the position to modulate the high-rate data.  Another well-known standardized under-sampling modulation technique for OCC is S2-PSK which uses a spatial modulation scheme \cite{ Nguyen2017}. A randomly sampled image can be fully demodulated by using S2-PSK due to the separation between the cameras. S2-PSK can demodulate randomly sampled images which allow it to be used for variable frame rate cameras. Recently, several other hybrid modulation schemes are standardized which include but not limited to pulse width modulation/pulse position modulation (PWM/PPM) \cite{Aoyama2015,Aoyama2015a}, camera multiple frequency shift keying (CM-FSK) \cite{Rajagopal2014,Lee2015,Hong2016}, camera on-off keying (C-OOK) \cite{Nguyen2016}, rolling shutter frequency shift keying (RS-FSK) \cite{Nguyen2016}, asynchronous quick link (A-QL) \cite{Nguyen2015c}, hidden asynchronous quick link (HA-QL) \cite{Nguyen2015c}, and variable transparent amplitude shape code (VTASC) \cite{Nguyen2015c}. 
\subsubsection{Non-Standardized Techniques}
In this section, we highlight numerous modulation schemes for OCC which are not standardized yet. 
In \cite{Penubaku2015}, a RGB color coding based modulation technique was used to transmit data from an LED display to the camera. The proposed technique in \cite{Penubaku2015} was tested for the transmission of different size images where the accuracy of the receiving camera decreases with the increase in image size. 
\footnotesize{
\begin{longtable}{|p{0.7cm}|p{2.0cm}|p{1.47cm}|p{4.5cm}|p{1.3cm}|p{1.3cm}|}
\caption{Comparison of modulation techniques for OCC.}\label{tabXX}\\
\hline
\hline
\textbf{Ref.}&\textbf{Data Rates}& \textbf{Distance (m)}&\textbf{Modulation Technique}&\textbf{System Design}&\textbf{Compl-exity}\\
\hline
\hline 
\cite{Liang2016}&	2.88 kbps&	0.1& Wavelength Division Multiplexing (WDM)  &LED2C&    {Low}\\ \hline
\cite{Lee2017}&	9 kbps&	0.1& Pulse Width Modulation (PWM)&LED2C&High\\ \hline
\cite{Liu2017}&	23 bps&	0.15&  UDPSOOK&LEDA2C &Low\\ \hline
\cite{Duque2016}&	0-1.55 kbps&	$0.05-0.4$& On-Off Keying (OOK)&LED2C&   {Low}\\ \hline
\cite{Cahyadi2016}&	1.28 kbps&	$0.05-0.2$&OOK&LEDA2C&   {Low}\\ \hline
\cite{Hu2015}&	5.2 kbps&	0.3&Color shift keying (CSK)&LED2C&   {Moderate}\\ \hline
\cite{Ji2014}&	15 bps &	0.1-0.3& Under-sampled Frequency Shift On-Off Keying (UFSOOK)&LEDA2C&   {Moderate}\\ \hline
\cite{Danakis2012}&	1-3.1 kbps&0.35&OOK&LED2C&   {Low}\\ \hline
\cite{Chen2014}&	0.24  kbps&	0.5&CSK&LEDA2C&   {Moderate}\\ \hline
\cite{Nguyen2015a} &	0.3-1.5 kbps &	0.5&PWM&LED2C&   {Low}\\ \hline
\cite{Chen2015}&	0.3 kbps&	0.5&OOK&LED2C&   {Low}\\ \hline
\cite{Takai2013}	&10,15,20 Mbps&	0.5,1,2&OOK&LEDA2C&   {Low}\\ \hline
\cite{Luo2015a}&0.15 kbps&0.6& Under-sampled Phase Shift On-Off Keying (UPSOOK) and WDM& LEDA2C&   {High}\\ \hline
\cite{Tian2016}&	95 kbps&	1.2&Color intensity modulation (CIM) and PAM&LEDA2C&   {High}\\ \hline
\cite{Huang2016}&	126.7 kbps&	1.4&CIM&LEDA2C&   {High}\\ \hline
\cite{Luo2015}&	0.25 kbps	&1.5& Under-Sampled Pulse Amplitude Modulation with Subcarrier Modulation (UPAMSM)&LED2C&   {Moderate}\\ \hline
\cite{Luo2016}&0.5 kbps	&1.5& Under-Sampled Quadrature Amplitude Modulation with Subcarrier Modulation (UQAMSM)&LED2C&   {High}\\ \hline
\cite{Hao2016a}&	0.62-1.35 kbps&	1.5-5.5 & Hybrid OOK-PWM&LED2C  LEDA2C&   {Moderate}\\ \hline
\cite{Kowalczyk2017}&	0.24 kbps&	4 &OOK&LEDA2C&   {Low}\\ \hline
\cite{Imai2016}&	84 kbps&	4&PWM&LED2C&   {Low}\\ \hline
\cite{Luo2014}&0.15 kbps& 12&UPSOOK& LED2C&   {Low} \\ \hline
\cite{Ong2016}	&2 bps	&25&OOK&LED2C&   {Low}\\ \hline
\cite{Ebihara2014}&1 kbps&	30& Spatially-Modulated Space-Time (SM-ST)&LEDA2C&   {High}\\ \hline
\cite{Song2015}& 0.1 kbps & 50&UPAMSM&LED2C&   {Moderate}\\ \hline
\cite{Ebihara2015}&	1 kbps&	40-210&  Layered Space-Time Code (L-STC) &LEDA2C&   {High}\\ \hline
\cite{Wang2015}&	0.12-0.96 Mbps&	0.12-0.24&  Color
Barcodes  &S2C&   {High}\\ \hline
\cite{Boubezari2016}&112.5 kbps &	0.2&CIM & S2C&   {High}\\ \hline
\cite{Du2016}&	317.3 kbps &	0.2&  CIM and Rateless codes   &S2C&   {High}\\ \hline
\cite{Wang2014}&	12.8 kbps&	0.5& CIM   &S2C&   {High}\\ \hline
\cite{Wang2016}&	240 kbps	&0.6&  Spatial-Temporal Complementary Frames (S-TCF)  &S2C&   {High}\\ \hline
\cite{Li2015}&	0.8-1.1 kbps&	0.3-1.5&  Pixel translucency modulation   &S2C&   {High}\\ \hline
\cite{Cahyadi2016a}&	11.52 kbps &	2 & CIM&S2C&   {High}\\ \hline
\cite{Hranilovic2006}&	1.344 Mbps&	2& Spatial Discrete Multitone (SDMT)    &S2C&   {High}\\ \hline
\cite{Perli2010}	&12 Mbps&	10&  OFDM  &S2C&   {High}\\ \hline
\hline
\end{longtable}}

A flicker-free, 2-PSK modulation technique was proposed in \cite{Nguyen2016a} for vehicular OCC which can achieve the data rate of 10 bps. Similarly, the authors in \cite{Kim2016}  experimentally demonstrated OCC system by using time shift light intensity modulation with the data rate in bits per second. Moreover, the authors in \cite{Imai2016a} used a polygon mirror with the image sensor to provide a data rate up to 120 bps. Pulse amplitude modulation (PAM) was expermentally tested in \cite{Krohn2018} which was able to achieve the data rate of 100 bps. Further improvement in data rate up to 250 bps was achieved in \cite{Song2015} and \cite{Luo2015} where the authors proposed under sampled PAM with subcarrier modulation. Correspondingly, to eliminate the phase uncertainity between the transmitter and the camera, a bulit-in gamma correction function was considered in \cite{Song2015} and \cite{Luo2015}. Consequently, in \cite{Arai2007}, the authors developed a fast Haar wavelet transform (2D FHWT) in which data was modulated on different spatial components based on the priority. The low-priority data was modulated onto the high-frequency components while the high-priority data was modulated on the low-frequency components. The effectiveness of the proposed scheme in \cite{Arai2007} was experimentally tested and verified. Recently, the authors in \cite{Luo2016} demonstrated an under-sampled quadrature amplitude modulation (QAM) subcarrier modulation technique for OCC. The proposed modulation technique in \cite{Luo2016} was experimentally tested which was able to achieve the data rate of 500 bps for a transmission distance of 1.5 m. Although the achievable distance and data rate of \cite{Luo2016} are low, it can be used for location-based services. 
A color shift keying technique (CSK) was used in \cite{Hu2015} to modulate the date by using different colors. The CSK technique was tested on different smart-phones where the maximum achievable data rate was 5.2 kbps with symbol error rate less than $10^{-3}$.
To improve the data rate for OCC, the authors in \cite{Tian2016} proposed a color intensity modulation (CIM) technique which achieves the data rate of 95 kbps over a transmission distance of 1.2 m. The authors in  \cite{Wang2015a} presented a code division multiple access (CDMA) like modulation technique with hierarchical frame structure called InFrame++. InFrame++ was able to embed messages into the video and can achieve the data rate of 360 kbps. 

In addition to the trend of improving the data rate and transmission distance, efforts are made to improve the system design for OCC, e.g.,   a spatial-angular modulation technique was developed in \cite{Han2018} by using pixelated MIMO systems to simplify the receiver design. Furthermore, an image sensor with the MIMO technique was used in \cite{Liang2016b} to reduce the inter-channel interference between the  RGB signals. Additionally, a hybrid modulation scheme of $M$-FSK and $N$-PSK was investigated in \cite{Hong2016} where $M$  is for multiple user access, and $N$ is for increasing the data rate. To cope with the background ambient light, the authors in \cite{Chen2015b} proposed a hybrid modulation scheme of spread spectrum and differential signaling. The authors in \cite{Le2016a}  proposed OCC by using the watermarks embedded in the images. The proposed idea in \cite{Le2016a} was tested by using an LCD as a transmitter and webcam as a receiver. A similar screen to camera communication technique (DisCo) was developed in \cite{Jo2016} which is robust to camera rotation, display size, occlusion, distortion, and blurring. The achievable data rate of DisCo was up to 1 kbps. Recently the authors in \cite{Yang2018n}  developed an OCC system called HYCACO which is based on spatial multiplexing and exploits the non-line of sight optical signals to improve the data rate up to 4.5 kbps. A LEDA2C communication system was proposed in \cite{Cahyadi2018} where near-infrared LEDs were used for transmission. The proposed model in \cite{Cahyadi2018} was experimentally tested that  achieved centimeter level localization accuracy and data rate of 6.72 kbps for 100 cm transmission distance. The authors in \cite{Chen2018} proposed a quadrichromatic LED-based OCC to improve both illumination and data rate. Color ratio modulation was used in \cite{Chen2018} which was able to achieve the data rate of 13.2 kbps at a distance of 2 cm. To improve the transmission rate and distance of S2C communications, the authors in \cite{Cahyadi2016a} used a dual-camera receiver with split shutter method. This method was able to achieve the data rate of 11.52 kbps at the transmission distance of 2 m.

The information carrying capabilities of OCC is also highly dependent on the type of transmitters and receivers. Based on the camera receiver, including modified and off-the-shelf cameras, the achievable data rate for OCC systems is diverse where a large body of demonstrations and experiments have been conducted. We have collected the experimental results of the literature and compared them in Table \ref{tabXX}.
\begin{table}[h]
\footnotesize
\centering
\caption{Comparison of coding schemes for OCC.}
\label{Tablecod}
\begin{tabular}{|p{1.20cm}|p{2.4cm}|p{3.9cm}|p{1.6cm}|p{1.6cm}|}
\hline
\hline
\textbf{Ref.}              & \textbf{System design}         & \textbf{Coding Scheme}  & \textbf{Data rate}&   {\textbf{Complex-ity}} \\ \hline \hline
\cite{Arai2007,Arai2008}& LEDA2C &Hierarchical Codes& 16, 128 kbps &    {Low}\\ \hline
\cite{Danakis2012}                     & LCD to camera  &  Manchester coding & -&    {Low}\\ \hline 
\cite{Takai2014}&	LEDA2C &BCH Code &10 Mbps&    {High}\\ \hline
\cite{Li2015}                      & S2C          & Alpha channel coding& - &   { Moderate}\\ \hline 
\cite{Hu2012}                      & LEDA2C         & RGB coding & 450 kbps &    {Moderate}    \\ \hline 
\cite{Nishimoto2011, Nishimoto2012}                      & LEDA2C           & Overlay coding & 0.25 kbps&    {Low}  \\ \hline 
\cite{Hu2012}& S2C	&Quick response (QR) codes & 450 kbps&    { Moderate}\\ \hline
\cite{Hu2013}&	S2C& Interframe Erasure Codes   &10 kbps&    {Low}\\ \hline
\cite{Ashok2014}& S2C	& QR codes & 31 Mbps &    { Moderate}\\ \hline
\cite{Wang2014a}                      & S2C   & Robust dynamic coding & 17 kbps&    { Moderate} \\ \hline 
\cite{Wang2015}                     & LEDA2C       & Rainbar coding & -  &    {High}   \\ \hline 
\cite{Du2016}                     & LEDA2C       & Rateless coding & 317 kbps &    { Moderate} \\ \hline
\cite{Nguyen2016} &	S2C &  Texture codes   & 22 kbps &   {Low}\\ \hline 
\cite{Chen2018a}                     & LEDA2C       & Alpha39 coding & 2.46 kbps&    {Low} \\ \hline
\cite{Shi2017}                     & LEDA2C       & Manchester coding & 4.32 kbps&    {Low} \\ \hline
\cite{Yang2017c}&	LEDA2C	&Raptor Codes& 5 kbps&    { Moderate}\\ \hline
\cite{Zhang2018b}                     & LEDA2C       & Reed-Solomon coding & - &    { Moderate} \\ \hline
\hline
\hline        
\end{tabular}
\end{table} 
\subsection{Error Detection Coding Schemes}
Error detection coding and decoding schemes for OCC has also been of great interest in the research community. In 
\cite{Danakis2012}, the authors have presented an encoder and decoder for OCC, based on  Manchester coding. An alpha channel based encoding technique is used in \cite{Li2015} for a screen to camera communications. The authors in  \cite{Li2015} have developed a prototype (HiLight) to demonstrate the efficiency and robustness of the proposed encoding scheme.
Similarly, the authors in \cite{Hu2012} have proposed an RGB coding technique to improve the transmission rate for OCC. A prototype based on \cite{Hu2012} was developed which was able to achieve the data rate of 450 kbps at a transmission distance of 1.5 meters. To further improve the data rate, the authors in \cite{Nishimoto2011, Nishimoto2012} have proposed an overlay coding technique in which original signals and inverted signals are transmitted for the long-range data while only original signals are transmitted for the short-range data.
The authors in \cite{Wang2014a} have improved the throughput of OCC by using robust dynamic barcode (RDCode). The RDCode enables a block structure for data packets which consists of error correction mechanisms at each block. Based on the block structure of RDCodes, it provides high-level transmission reliability. An improved color barcode based coding scheme (Rainbar) has been proposed in \cite{Wang2015} which allows accurate extraction of code and provides flexible synchronization which is essential for robust localization.  A rateless channel coding approach (SoftLight) has been proposed in \cite{Du2016} which considers the optical channel characteristics and adapts the data rate according to the qualities of the links. The authors in \cite{Chen2018a} have developed a LEDA2C communication system which uses the Alpha39 code and OOK modulation with the achievable data rate of 2.46 kbps for mobile phone payments. Recently, Manchester coding with OOK modulation has been used in \cite{Shi2017} for OCC to achieve the data rate of 4.32 kbps. Reed-Solomon codes were proposed in \cite{Zhang2018b} for LEDA2C communication to detect ultra long IDs of LEDs for improving the localization accuracy and reducing the complexity. Table \ref{Tablecod} summarizes the characteristics of different coding schemes for OCC.
\subsection{Synchronization}
Synchronization is an important issue for seamless communication between the transmitter and the receiver. In the case of OCC systems, the transmitter and receiver have different characteristics (LEDs have high data rate while the cameras have a low frame rate) which make the synchronization more challenging. In \cite{Hu2013}, the authors have proposed a synchronization method for LCDs and smartphone camera for OCC communication. A frame generator called LightSync was used for the LCD which modifies the original frames by adopting a linear code. The authors in \cite{Hu2013} were able to recover the lost frames and synchronize the LCD to the smartphone camera. The authors in \cite{Rajagopal2014} proposed to simultaneously transmit a common preamble for the data packet to achieve synchronization. In their proposed synchronization scheme, they have shown that periodically transmitting the packets allows synchronization between the receiver and the infrastructure.
The effect of synchronization between the transmitter and the receiver has been studied in the form of inter-frame packet loss for LEDA2C communication where an error correction coding scheme was proposed to solve this problem \cite{Hu2015}. Recently, a flexible synchronization method of OCC systems has been proposed in \cite{ Shiraki2017} which does not require synchronization between the LEDs and the receiving camera. Moreover, the authors in \cite{Akiyama2016} have used the smartphone microphones and camera to receive the acoustic signals and to synchronize time respectively. The transmitter in \cite{Akiyama2016} consists of loudspeakers and an LED. Recently, the authors in \cite{Kwon2017} have proposed a synchronization method for the LEDA2C communication system which ensures effective synchronization to maintain the color uniformity for even a camera with the frame rate of 30 frames per second. The authors have further improved the time synchronization method in \cite{Kwon2018} where color independent characteristics of visual MIMO have been exploited. Table \ref{Tablevlmod} summarizes the different synchronization methods developed for OCC. 
\begin{table}[h]
\footnotesize
\centering
\caption{Comparison of Synchronization methods for OCC.}
\label{Tablevlmod}
\begin{tabular}{|p{1.30cm}|p{2.6cm}|p{6.4cm}|}
\hline
\hline
\textbf{Ref.}              & \textbf{System design}         & \textbf{Synchronization}  \\ \hline \hline
\cite{Hu2013}                     & LCD to camera  &  Using linear codes \\ \hline 
\cite{Rajagopal2014}                      & LEDA2C          & Using a common preamble \\ \hline 
\cite{Hu2015}                      & LEDA2C         & Using error correction codes     \\ \hline 
\cite{ Shiraki2017}                      & LEDA2C           & Not required \\ \hline 
\cite{Akiyama2016}                      & LED2C   & Using time of arrival measurements \\ \hline 
\cite{Kwon2017}\cite{Kwon2018}                     & LEDA2C       & Using the standard deviation of light intensity      \\ \hline 
\hline
\hline        
\end{tabular}
\end{table} 
\subsection{Summary and Insights}
In this section, we have covered the physical layer aspects of OCC which include channel characterization, modulation, coding, and synchronization. Two different channel models have been presented for a single LED and LED arrays respectively. The received signal in OCC is affected by different kinds of noise such as shot noise, read noise, and non-uniform photo response noise. Also, the gain of received signal increases in the case of using multiple LEDs for the transmission. Furthermore, different modulation schemes are classified based on the IEEE standardization. The literature on modulation schemes for OCC is rich where the research trend mainly focuses on developing novel modulation techniques to improve the data rate and transmission range. In addition to the modulation techniques,  we have surveyed the existing coding schemes developed for OCC. Finally,  we have highlighted the synchronization problem of OCC  and presented the current synchronization schemes.

\section{Signal Processing Techniques for OCC}
This section covers the signal processing techniques for OCC. Mainly, we focus on the implementation of OCC by using color-based modulation schemes such as color shift keying (CSK), color and intensity shift keying (CISK), color intensity modulation (CIM), and CIM- MIMO.
\begin{figure*}
  \centering
  \captionsetup{justification=centering}
  \includegraphics[scale=0.5]{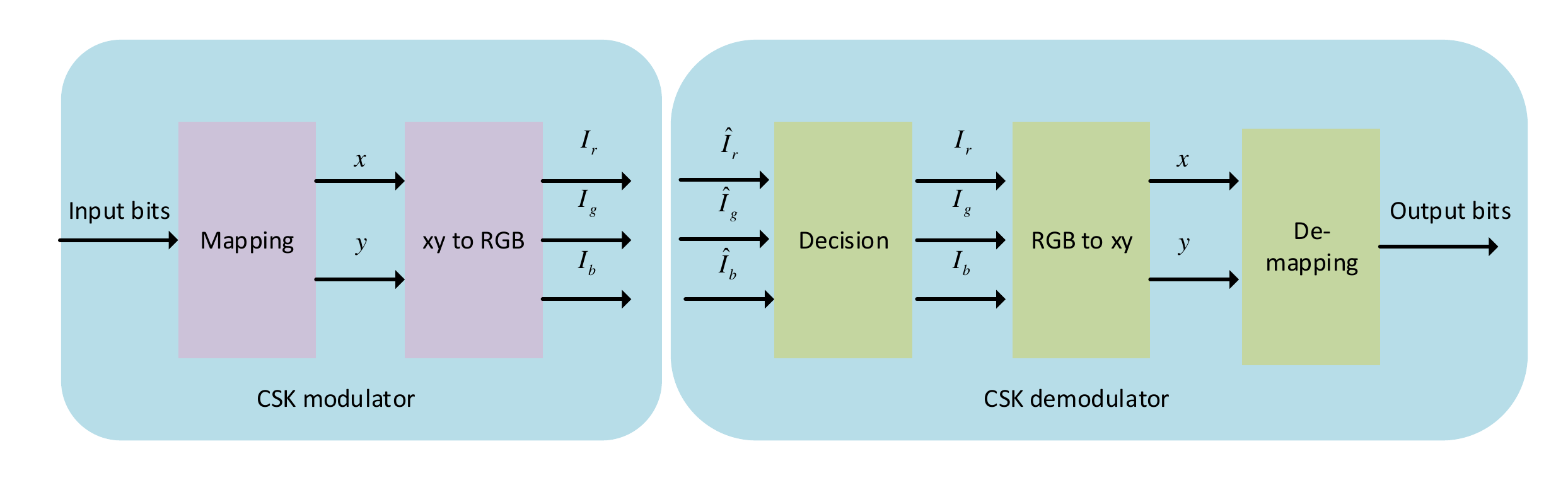}\\
  \caption{CSK modulation and demodulation.}\label{fig:csk}
\end{figure*}

\subsection{Color Shift Keying:}
The IEEE 802.15.7 standard proposed CSK modulation for OCC because the receiver in OCC is an image sensor or a camera that works as a color filter.  CSK technique requires the use of three different color LEDs (red, green, and blue) for modulation. The information is modulated by mixing the intensities of the three colors based on the color space chromaticity diagram. Based on different wavelengths of the LEDs, the input data is mapped into $M$-ary symbols in the CIE color space \cite{commission1924} as shown in Fig. \ref{fig:csk}.  The chromaticity values are then transformed into the intensities values of $I_r, I_g, I_b$ which relates to symbols $x,y$ as follows
\begin{equation}
x = I_r \cdot x_r + I_g \cdot x_g + I_b \cdot x_b,
\end{equation}
\begin{equation}
y = I_r \cdot y_r + I_g \cdot y_g + I_b \cdot y_b,
\end{equation}
where $(x_r, y_r)$, $(x_g,y_g)$, and $(x_b,y_b)$ are the chromaticity values of red, green, and blue colors respectively. Hence, the resultant signal space $S$ is function of the intensities values of $I_r, I_g, I_b$. At the receiver (image sensor/camera), the three color intensities are estimated by using the concept of minimum Euclidean distance given as
\begin{equation}
\bar{S} = \arg \min_{S \in \mathcal{M}}\parallel\hat{S} - S\parallel
\end{equation}
where $\mathcal{M}$ shows the set of $M$-ary CSK modulation, $\hat{S}$ is the received symbol, and $\parallel\cdot \parallel$ is the Euclidean norm. Based on the above analysis, the bit error rate (BER) for CSK modulation in AWGN is obtained as follows \cite{Tang2018}
\begin{equation}
\text{BER} = \frac{1}{\log_2(M)}\sum_{i=0}^{M-1}\sum_{j=0,j\neq i}^{M-1} d(S_i,S_j)P(S_j/S_i)P(S_i),
\end{equation}
where $d(S_i,S_j)$ is the Hamming distance between symbol $S_i$ and $S_j$,  $P(S_j/S_i)$ is the transition probability, and $P(S_i)$ is the a prior probability of $S_i$. Recently, the CSK is combined with undersampled pulse amplitude modulation (UPAM) to provide extra degree of freedoms in both intensity and color domain. Combining CSK with UPAM (CISK) yields better spectral efficiency as compared to the conventional CSK technique \cite{Chen2019}.

\subsection{Color Intensity Modulation:}
The CSK modulation technique when combined with multi-level PAM, results in color intensity modulation (CIM). In CIM, the set of color pixels are partitioned into sub-blocks, resulting in a MIMO configuration. Hence, an extra degree of freedom (space) is achieved in CIM-MIMO. In the CIM-MIMO setup, current from the LEDs is mapped into the corresponding color and intensity level. For example, in case of three colors and $I$ intensity levels, the number of symbols $M$ is equal to $I^3$, where each symbol $S$ is the function of intensity levels, i.e., $S = [I_r, I_g, I_b]$. The symbols are transmitted over the wireless channel and received by the image sensor/camera, where an $N \times N$ channel matrix is calculated as follows
\begin{equation}
\mathbf{H} = \begin{bmatrix}
\mathbf{H}_{1,1} & \cdots & \mathbf{H}_{1,N}\\
\vdots & \ddots & \vdots \\
\mathbf{H}_{N,1}& \cdots & \mathbf{H}_{N,N}
\end{bmatrix},
\end{equation}
where $\mathbf{H}_{i,j}$ represents the channel gain matrix between the $i$-th transmitter LED and $j$-th receiver. Based on the above channel matrix, the received signal $Y_i$ at the $i$-th receiver in the presence of AWGN channel is given as
\begin{equation}
Y_i = \mathbf{H}_{i,i}X_i + \sum_{i\neq j}\mathbf{H}_{i,j}X_j + Z_i,~~ i,j \in 1,2,...N
\end{equation}
where $X_i$ is the transmitted symbol and $Z_i$ is the Gaussian noise. Given all the observations $\mathbf{Y}_i,~i = 1,2,...N$, and the channel matrix $\mathbf{H}$, symbols at the receiver can be estimated by using the maximum likelihood estimator as follows \cite{Huang2016}
\begin{align}
\hat{\mathbf{X}}_i &= \arg \max_{\mathbf{X}_i} p(\mathbf{Y}|\mathbf{X}_i,\mathbf{H}) \nonumber \\
 &= \arg \min_{\mathbf{X}_i}\parallel \mathbf{Y} - \mathbf{H} \mathbf{X}_i\parallel_F,
\end{align}
where $\parallel \cdot \parallel_F$ is Frobenius norm. Based on the above estimator, the symbol error probability $P_e$ is obtained in \cite{Singh2014} as follows
\begin{equation}
P_e = \frac{1}{M}\sum_{k=1}^M\left\lbrace  N_{ik}Q\left(\sqrt{\frac{d^2}{2 N_0}}\right)\right\rbrace
\end{equation}
where $N_{ik}$ represents the number of nearest neighbors of the $k$-th symbol, $d$ is the minimum distance between symbols, $N_0$ is the one-sided power spectral density, and $Q(\cdot)$ is the tail probability of normal distribution.

\section{Use cases of OCC}
This section briefly covers major applications of OCC that include localization, navigation, motion capture, and intelligent transportation systems.

\subsection{OCC-based Localization and Navigation}
Location-based services (LBS) require to develop positioning systems which are able to estimate the user location \cite{Kaplan2006}. The most widely used positioning system in many applications such as navigation, tracking, surveillance, and public safety is global positioning system (GPS). However, in unfriendly environments such as indoor and urban canyons, the signals from the GPS satellites are obstructed which results in limited coverage and inaccurate results \cite{Bulusu2000}. Therefore, a number of indoor positioning systems (IPS) have been developed in the past decade to overcome the limitations of GPS. These IPS are based on different wireless technologies such as radio frequency identification (RFID) \cite{Jimenez2012, Yang2014, saeed2016efficient, saeed2016robust}, Wi-Fi \cite{He2016}, Bluetooth \cite{Hossain2007}, acoustic \cite{Kantarci2011}, infrared \cite{Chen2010}, and Zigbee \cite{Fang2012}. Among all these technologies Wi-Fi and Bluetooth based IPS have been predominantly utilized and embedded in the current smart devices.

Recently, research on visible light communication (VLC) has gained much interest in both academia and industry due to the numerous applications of VLC systems \cite{Sevincer2013, Karunatilaka2015,Saha2015,Koonen2018}. There are many applications of VLC systems, including, but not limited to, vehicular communication \cite{Dimian2017}, underwater communication \cite{Saeed2018, Saeedtmc, Saeedtcom, Nasir2018twc}, and IPS \cite{Hyun2015}. A number of localization algorithms have been proposed over the past few years for VLC and OCC based IPS \cite{Kuo2014, Hyun2015,Hassan2015, Trong2016, Luo2017n}. Experiments on VLC based IPS have shown that it can provide a sub-meter level accuracy while the accuracy of Wi-Fi and Bluetooth based IPS have an accuracy of (1-7 m) and (2-5 m) respectively \cite{Hassan2015}. Nevertheless, (\textit{Visible light beacon system}) developed in \cite{Yoshizawa2016} can provide millimeter level of accuracy.

\begin{figure}[h]
  \centering
  \includegraphics[scale=0.65]{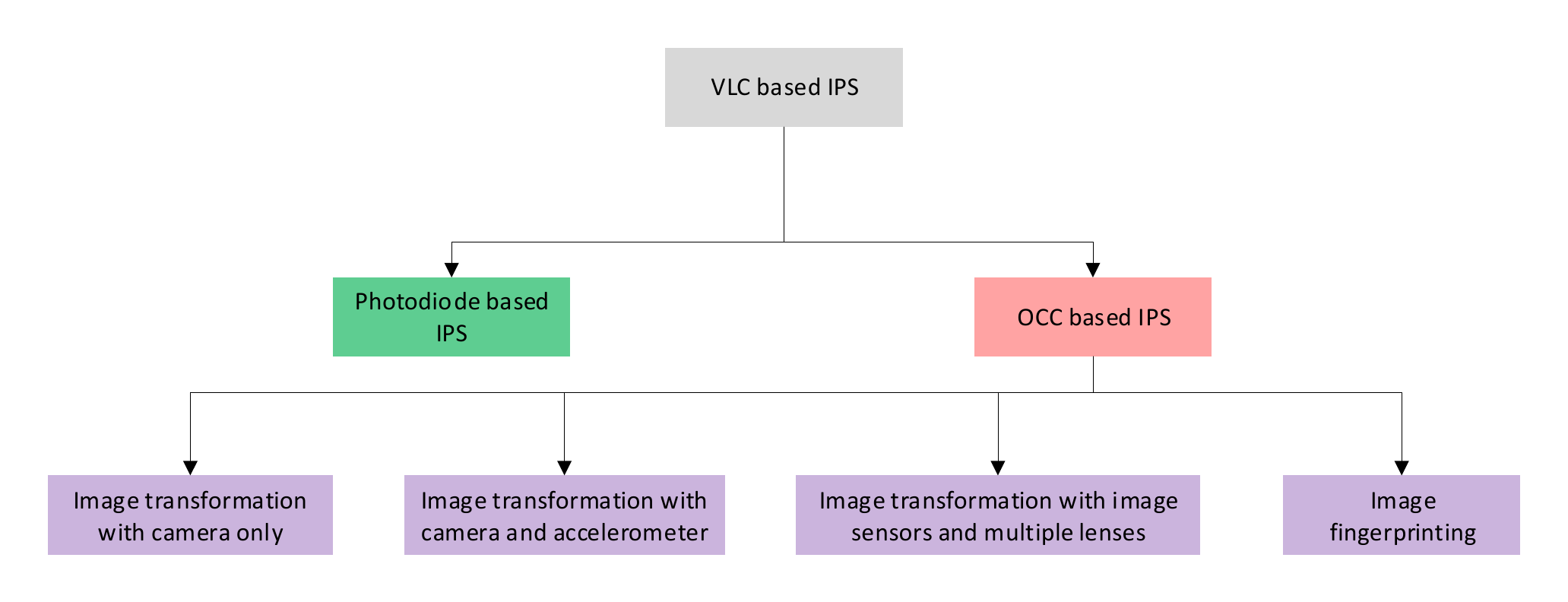}\\
  \caption{Classification of VLC based IPS.}\label{fig:IPSclass}
\end{figure}
In the past few years, a good number of surveys have been published on VLC and VLC based IPS \cite{Sevincer2013, Karunatilaka2015,Saha2015,Koonen2018}. For instance, the authors in \cite{Pathak2015} have studied the characteristics of VLC, medium access techniques, channel properties, system design, and applications of VLC. The literature on IPS by using VLC has been presented in \cite{Hassan2015}. The authors in \cite{Zhang2017a} have compared the performance of three different VLC based IPS. A comprehensive survey on VLC based IPS have been presented in \cite{Trong2016} which classifies the different IPS based on the localization technique used and its accuracy. The VLC based IPS can be broadly classified into PD based systems and OCC based systems as shown in Fig.~ \ref{fig:IPSclass}. In this paper, we are mainly focusing on the OCC based IPS.
\subsubsection{OCC Based IPS}
\begin{figure}[h]
  \centering
  \includegraphics[scale=0.5]{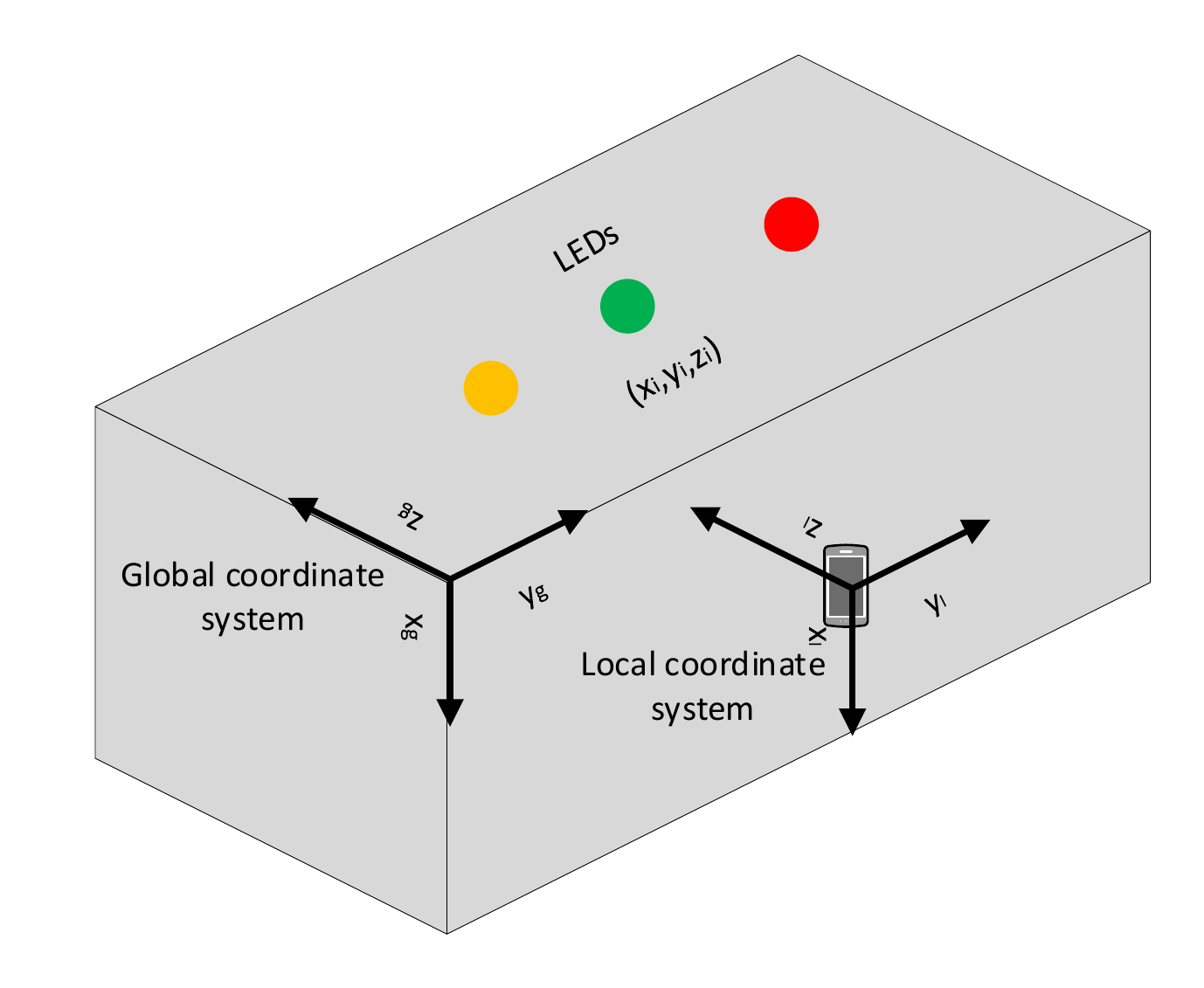}\\
  \caption{Global and local coordinate systems.}\label{fig:IPSit}
\end{figure}

OCC-based IPS consists of visible LEDs as transmitters (beacons) while the receiver consists of a cameras or image sensors which collect the images from the LED sources. Location of the LED in the taken image defines the location of the LED with respect to the lens. The global position estimate of the LED is determined by taking consecutive images of the LED. The minimum number of required images to estimate the position of LED depends on the transmission code length and switching frequency of the LED. For example, the images taken by a camera from an LED which uses OOK modulation, are examined by either color balance, edge detection, or pixel detection method to determine the ON/OFF state of the LED.
Once the state of LED is determined, the global position of LED can be estimated by using the code sequence. Finally, the image transformation techniques are used to determine the global position and orientation of the camera. Different image transformation techniques are used for the OCC based final position estimation as shown in Fig~\ref{fig:IPSit}.
\paragraph{Image Transformations by Using Camera Only}
It is one of the most widely used techniques for OCC based IPS where the LED transmits their three-dimensional (3D) position coordinates to the camera \cite{Liu2003, Yoshino2008,Kuo2014,Han2016,lumicast2016, Lin2017}. 
The camera demodulates the signals from the LEDs and determines the 3D coordinates of the LEDs and itself. Fig.~\ref{fig:IPSit} explains the image transformation based IPS by using OCC. In Fig.~\ref{fig:IPSit} $\{x_l,y_l,z_l\}$ represents the local coordinate system of camera while $\{x_i,y_i,z_i\}$ represents the global coordinates of the $i$-th LED. The global coordinates and orientation of the camera are denoted by $\{x_g,y_g,z_g\}$ and $\{\alpha,\beta,\gamma\}$
respectively. The transformation factors $\boldsymbol{T}_\alpha$, $\boldsymbol{T}_\beta$, $\boldsymbol{T}_\gamma$ for the three-dimensional coordinates are given as follows
\begin{equation}
\boldsymbol{T}_\alpha = \begin{bmatrix}
1 & 0 & 0\\
0 & \cos(\alpha) & -\sin(\alpha)\\
0 & \sin(\alpha) & \cos(\alpha)\\
\end{bmatrix},
\end{equation}
\begin{equation}
\boldsymbol{T}_\beta = \begin{bmatrix}
\cos(\beta) & 0 & \sin(\beta)\\
0 & 1 & 0\\
-\sin(\beta) & 0 & \cos(\beta)\\
\end{bmatrix},
\end{equation}
and
\begin{equation}
\boldsymbol{T}_\gamma = \begin{bmatrix}
\cos(\gamma) & -\sin(\gamma) & 0\\
\sin(\gamma) & \cos(\gamma) & 0\\
0 & 0 & 1\\
\end{bmatrix}.
\end{equation}
The transformation factors $\boldsymbol{T}_\alpha$, $\boldsymbol{T}_\beta$, $\boldsymbol{T}_\gamma$ are used for the rotation along $x$-axis, $y$-axis, and $z$-axis respectively \cite{Hijikata2009}.  Consider vector $\boldsymbol{l}=[x_l-x_i, y_l-y_i, z_l-z_i]^T$ from the origin of the local coordinate system to the $i$-th LED which is parallel to the direction vector $\boldsymbol{\gamma} = [a_i,1,b_i]^T$ in the image coordinate system. Hence, the following condition is satisfied
\begin{equation}\label{eq: px}
[a_i,1,b_i]^T \parallel R^{-1} [x_l-x_i, y_l-y_i, z_l-z_i]^T,
\end{equation}
where $R = \boldsymbol{T}_\alpha \times \boldsymbol{T}_\beta \times \boldsymbol{T}_\gamma$. From \eqref{eq: px} $a_i$ and $b_i$ can be expressed as 
\begin{eqnarray}\label{eq: ip1}
a_i &=& \frac{1}{Y}\Bigg(x_l-x_i)(\cos \gamma \cos \beta-\sin \gamma \sin \beta \sin \alpha)\nonumber \\
& & \Bigg. + (y_l-y_i)(\sin \gamma \sin \beta-\cos \gamma \cos \beta \cos \alpha)\nonumber \\
& & \Bigg. - (z_l-z_i)\cos \gamma \cos \beta \Bigg),
\end{eqnarray}
 and
\begin{eqnarray}\label{eq: ip2}
b_i &=& \frac{1}{Y}\Bigg((x_l-x_i)(\cos \gamma \sin \beta-\sin \gamma \sin \alpha \cos \beta)\nonumber \\
& & \Bigg. + (y_l-y_i)(\sin \gamma \sin \beta-\cos \gamma \sin \alpha \cos \beta)\nonumber \\
& & \Bigg. - (z_l-z_i)\cos \gamma \cos \beta \Bigg),
\end{eqnarray}
respectively, where $Y = -(x_l-x_i)\sin \gamma \cos \alpha+ (y_l-y_i) \cos \gamma \cos \alpha + (z_l-z_i) \sin \alpha$. Note that there are six unknown parameters in \eqref{eq: ip1} and \eqref{eq: ip2}, i.e., $\{x_l,y_l,z_l, \alpha, \beta, \gamma\}$. All these unknown parameters can be found from \eqref{eq: ip1} and \eqref{eq: ip2} by using non-linear least square estimation.

In \cite{Nakazawa2013} the authors have used a camera with fish eye lens to capture the images from the LED lights and utilized the image processing tools to estimate the global coordinates of the camera. The authors in \cite{Nakazawa2013} have experimentally tested their proposed system at Niigeta university, Japan and achieved an accuracy of 10 cm. In \cite{Ifthekhar2014} the authors have used photogrammetry to map the 3D environment into  2D image. Neural networks were used in \cite{Ifthekhar2014} to estimate the distance to the LEDs. Recently, in \cite{Hossan2018} the authors introduced the OCC-based localization for vehicles where photogrammetry is used to estimate the distance between the vehicles. Although such kind of systems can measure both the location and orientation of the camera but they require complex and costly image processing techniques. Therefore, the authors in \cite{Kim2013} have proposed a more fast and accurate OCC based IPS which utilizes the intensity and carrier allocation methods. The authors in \cite{Kim2016, Wang2018loc} have also proposed an image processing based IPS by using two LED transmitters and a receiving camera. However, detection of images from LEDs by using smart phone camera and further processing them for localization is a challenging task because the hardware of smart phones is not designed to support OCC based positioning. Therefore, the authors in \cite{Li2017a}  have introduced the use of Miller codes to robustly decode the optical signals for OCC-IPS. Recently, the authors in \cite{Zhu2017,Zhu2018} have experimentally evaluated an angle difference of arrival (ADOA) based ranging for OCC-IPS where the authors have claimed an accuracy of 3.20 cm.  
\paragraph{Image Transformation by Using Camera and Accelerometer}
The complexity of the image transformations methods can be reduced by using the accelerometer \cite{ Yoshino2008}. Once the accelerometer is calibrated it can provide the coordinates information of the user device along each direction and facilitate to find the orientation angles of the camera. Hence, this approach does not rely fully on the image transformation methods to compute the orientation angles. Once the orientation angles are determined, the location of the user device is estimated by using the image transformation methods. In \cite{Jung2011} the authors have proposed OCC based IPS which uses the difference in angle of arrival to compute the orientation angles, however, the authors in \cite{Jung2011} assume that the receiver gain profile of the device is known prior which is not practical. Similarly, in \cite{Yasir14,Liu15} the authors have presented an OCC-IPS based on phase difference of the received signals from different transmitters in the line of sight (LoS) condition. However, in the real indoor environment, the received signals may be reflected from different obstacles resulting in noisy ranges which can severely degrade the localization accuracy. Therefore, recently, in \cite{Huang16} the authors have proposed an OCC-IPS which computes the ranges from the accelerometer/gyrometer measurements and light intensity of the transmitters.  
\begin{table}[h]
\footnotesize
\centering
\caption{Comparison of different Localization and navigation systems based on OCC.}
\label{Tableoccloc}
\begin{tabular}{|p{1.0cm}|p{4.0cm}|p{3.0cm}|p{2.0cm}|}
\hline
\hline
\textbf{Ref.}              & \textbf{Receiver type}         & \textbf{Accuracy} & \textbf{Processing time} \\ \hline \hline
\cite{Rajagopal2014}                     & Camera (3 Megapixel)  &  room-level & 528 ms \\ \hline
\cite{Li2014epsilon}                     & Camera  &  0.4 - 0.8 m & 0.7 s \\ \hline 
\cite{Kuo2014}                     & Camera (33 Megapixel)  &  7 cm & 0.9 s \\ \hline
\cite{Yasir14}                     & Camera   &  0.25 m & - \\ \hline
\cite{Yang2018pixel}                     & Camera (8 Megapixel)  &  30 cm & 1.2 s \\ \hline 
\cite{lumicast2016}                     & Camera  &  10 cm & 0.1 s \\ \hline 
\cite{Yoshino2008}                     & Image sensor  &  1.5 m & - \\ \hline
\cite{Han2016}                     & Image sensor  & 5 cm & - \\ \hline
\cite{Lin2017}                     & Camera  & 6.6 cm & - \\ \hline
\cite{Nakazawa2013}                     & Camera  & 10 cm & - \\ \hline
\cite{Hossan2018}                     & Camera  & 10 cm & 10 s \\ \hline
\cite{Kim2013}                     & Camera  & 2.4 cm & - \\ \hline
\cite{Wang2018loc}                     & Camera  & sub-meter-level & 0.58 s \\ \hline
\cite{Zhu2017}                     & Camera  & centimeter-level & 0.2 s \\ \hline
\cite{Zhu2018}                     & Image sensor  & 3.20 - 14.66 cm & 0.36 - 0.001 s \\ \hline
\hline
\hline        
\end{tabular}
\end{table}  
\paragraph{Image Transformation by Using Image Sensors and Multiple Lenses}
The other way to reduce the complexity of geometrical relationships between the transmitter LEDs and the receiver is to use multiple image sensors where a lens is placed in front of each image sensor \cite{Rahman2011,Rahman2011a}. The estimated location of the receiver is the midpoint of the line connecting the two lenses. This approach requires at least three non-collinear LEDs and the receiver position is estimated from the position difference of the LEDs. The use of multiple image sensors and lenses can reduce the complexity of the localization method but increases the overall cost of the system. Recently, the authors in \cite{Zhang2016a} have formulated the problem of OCC-IPS as a multi-objective non-convex optimization where the information from image sensor and motion sensor were fused. Singular value decomposition based algorithm was proposed in \cite{Zhang2016a} to fuse the information from the sensors.    {Nevertheless, using multiple cameras operating simultaneously at the same frequency introduces interference between the cameras that can cause a high ranging error. One solution to avoid the inter-camera interference is to employ time division multiplexing where each camera operates at a different time period. Nonetheless, this will require perfect synchronization between the cameras leading to extra cost. Frequency division multiplexing can also be used to avoid inter-camera interference by operating the cameras at different frequencies. However, due to the limited number of available frequencies, only a few cameras can work at the same time. Methods based on pseudo-noise cancellation are also used in literature to avoid the inter-camera interference \cite{Buttgen2007, Whyte2010}. These techniques separate the reflections from the nearby cameras; yet, they require the cameras to be upgraded such that the cameras can operate on sinusoidal modulation. Recently, in \cite{Li2015I}, the authors proposed a least square estimation method to mitigate the inter-camera interference where the interference signal was proved to be a wide-sense stationary stochastic process. }

\subsubsection{Summary and Insights}
There are quite a few survey papers on VLC-based IPS, however, OCC-based IPS have not been the focus of those surveys. Therefore, in this section, we have surveyed and classified the literature on OCC-based IPS as shown in Fig.~\ref{fig:IPSclass}. All of the OCC-based IPS require image processing techniques to estimate the location of the user where the images from the LEDs are captured by the camera and further processed to find the location of the user. These image transformation based methods are very complex and thus are not suited for hardware of the smart phones. 
In Table \ref{Tableoccloc} we present well-known localization systems based on OCC.

\subsection{OCC-based Motion Capture}
The idea of motion capture was first introduced by Eadweard Muybridge back in 1887 where the movement of birds and animals was studied. The motion of the subject was estimated by taking the photographs at discrete time intervals. Similarly, an experiment (moving light display) was conducted by Johansson in 1973 to visualize the biological motion of different subjects \cite{Johansson1973}. These experiments lead the research of motion capture which has grown up to support a large number of applications such as clinical studies, surveillance, character recognition, gesture recognition, and computer games. 
\begin{figure}[h]
  \centering
  \includegraphics[scale=0.75]{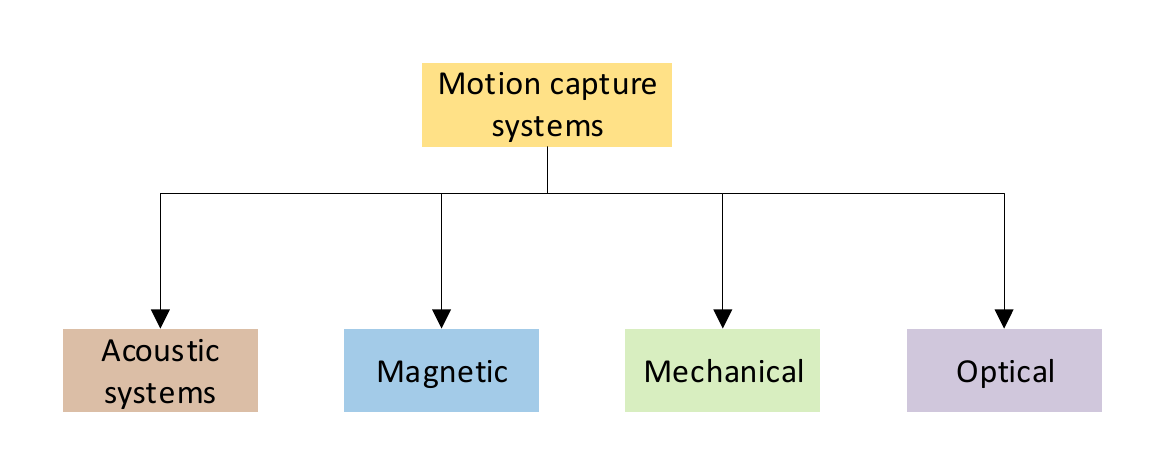}\\
  \caption{Classification of motion capture systems.}\label{fig:motion}
\end{figure}

The analysis of a visual event where the movement of a subject is represented mathematically is known as motion capture. For example, the task of observing the human actions which require movement can be accomplished by putting sensors on the body parts of the human where the movement of the body is estimated from the movement of the sensors. Motion capture has gained much interest recently, and the papers published in this research area has grown exponentially. The existing motion capture systems can be classified into four classes, i.e., acoustical, mechanical, magnetic, and optical systems (see Fig.~\ref{fig:motion}). As this survey is focused on OCC, we will restrict our discussion to the optical systems.

The idea of motion capture has recently been brought in as an added service to the VLCs, where the primary functions are communication and illumination \cite{Sewaiwar2015}. Motion capture schemes are considered to be viable and attractive to control different smart devices. Numerous VLC based motion capture schemes have been developed \cite{Sewaiwar2015,Ionescu2013,TIWARI2015,LALITHAMANI2016} for various applications. 
However, the literature on motion capture schemes for OCC based systems is not rich. In \cite{Seaman2012} and \cite{Solberg2016}, the authors have evaluated the performance of optical and inertial motion capture systems where the optical system provided better accuracy (sub-millimeter). The optical motion capture system developed in \cite{Seaman2012} and \cite{Solberg2016} consists of cameras which emit infrared light and reflective sensors placed on the subject which reflect back the light to the cameras. Indeed, there are a large number of commercial optical motion capture systems developed which are based on the concept of camera transmitting the infrared light and receiving the reflected light  \cite{ma2018,Jackson2018,Stt2018,opti2018,meta2018}. Although optical motion capture systems based on the reflective technologies provide real time tracking of the movement but their accuracy is less than LEDs based technology  becasue LEDs have unique ID. LEDs based active optical motion capture system is developed in \cite{richard2018}, where each LED flashes at a particular frequency which makes it easier for tracking. In order to explain the working principle of LEDs based active motion capture, the receiving camera is considered as a pin-hole camera where the light enters the camera aperture and is projected onto the image plane. Consider that $\mathbf{p} = \{x,y,z\}$ is a vector of camera coordinates of the subject, then the point in the image where $\mathbf{p}$ projects is given by $\hat{\mathbf{p}}=\mathbf{T}\mathbf{p}$, where $\mathbf{T}$ is the projection matrix from real-world subject to the camera image plane. The projection matrix $\mathbf{T}$ is calculated based on the focal length $f$ of the camera given as
\begin{equation}\label{eq: T}
\mathbf{T} = \begin{bmatrix}
f & 0 & 0 & 0\\
0 & f & 0 & 0\\
0 & 0 & 1 & 0\\
\end{bmatrix}.
\end{equation} 

Undoubtedly, $\mathbf{T}$ in {\color{red}\eqref{eq: T}} represents the simplest form of the projection matrix. However, in reality the intrinsic and extrinsic parameters (calibration) is also required. The interested readers are referred to \cite{richard2018, Shin2009, Shin2012} for calibration of the camera in optical motion capture systems. Nevertheless, optical motion capture systems based on high-resolution cameras can be costly whereas the smart-phones cameras have improved significantly which supports the applications of motion capture. Shivani \textit{et al} have proposed a smart-phone based motion capture system where they have used an $8 \times 8$ array of LEDs for the transmission, and a smart-phone camera is used as a receiver  \cite{Teli2017}. The screen of the smart-phone was divided into four quadrants where the proposed scheme was able to identify three different shapes, a line, L shape, and a circle.
The authors have experimentally evaluated the OCC based motion capture system where the accuracy of 96 \% is achieved with the data rate of 1216 bps. To further improve the detection and data rate of optical motion capture systems, the authors in\cite{Shivani2018} have proposed a neural network based approach. The experimental setup consists of an array of $8 \times 8$ LEDs as a transmitter and a smart-phone camera as a receiver. The proposed algorithm considers the neurons which were trained with the motion of the subject as well as with the data samples. The proposed scheme was able to detect accurately the shapes of a line, curve, circle, and semicircle and was also able to improve the data rate up to 3.759 kbps.

\paragraph*{Summary and Insights} In this section, we have briefly introduced the motion capture systems which can be implemented using different technologies. 
    {The literature on conventional motion capture systems is rich; however, due to the scope of the survey, we have restricted our discussions only to the optical motion capture systems.} Unlike the traditional reflective systems, in optical motion capture systems, the cameras are responsible for collecting the images of the subject and estimate the movement. Although the optical motion capture systems are more adaptable and provide better accuracy, it requires accurate calibration of the cameras. Additionally, the correspondence in space and time domain is a significant challenge of optical motion capture systems which is still under research.

\subsection{OCC-based Intelligent Transportation Systems}

LEDs have both capabilities of providing light and communication with low power consumption and low heat generation. Hence, these features make them more suited to provide intelligent transportation system (ITS).
LEDs-based ITS can help in regulating the traffic and avoiding traffic accidents. For example, LEDs-based traffic lights can broadcast driving assistance signals to the vehicles (I2V) and the LEDs-based vehicle brake lights can be used to transmit warning signals to a nearby vehicle or the infrastructure (V2V/V2I).  Moreover, OCC is also applicable for the ITS which enables both V2V and V2I communication. The cameras in vehicles can receive the optical signals from the traffic lights, nearby vehicle headlights, brake lights, and signage. The notable feature of OCC-based ITS is that the cameras provide spatial separation where only the meaningful pixels are focused and the noisy pixels are discarded.
\begin{table}[h]
\footnotesize
\centering
\caption{Summary of the literature for OCC-based ITS}
\label{Tableitocc}
\begin{tabular}{|p{0.6cm}|p{2.3cm}|p{1.0cm}|p{2.5cm}|p{4.5cm}|}
\hline
\hline
\textbf{Ref.}              & \textbf{Scenario}         & \textbf{Data rate} & \textbf{Transmission distance} & \textbf{Modulation/coding}  \\ \hline
\cite{Takai2013}                     & I2V static  &  10-20 Mbps & 2 m & OFDM, QAM/Manchester coding \\ \hline 
\cite{Takai2014}                     & V2V mobile  &  10 Mbps & 2 m & OFDM, QAM/Manchester coding \\ \hline
\cite{Yamazato2014}                     & V2I/V2V mobile  &  10 Mbps & 30-70 m & OFDM, QAM/Manchester coding \\ \hline
\cite{Yamazato2014}                     & I2V mobile  &  10 Mbps & 30-70 m & OFDM, QAM/Manchester coding \\ \hline
\cite{Yamazato2016}                     & I2V static  &  - & 30-60 m & - \\
\hline
\cite{Teli2018,TELI2018241}                     & V2V static  &  6.912 kbps & 175 cm & OOK \\ \hline
\hline      
\end{tabular}
\end{table}

OCC-based ITS requires high-speed capturing and extraction of the data to ensure flicker-free  LEDs in the vehicle. This issue was addressed in \cite{roberts2014automotive} with a high-speed camera (1000 fps). Nevertheless, the authors in \cite{Teli2018,TELI2018241} used a more realistic camera with 60 fps of speed where the idea of selective capturing was introduced.
The field tests in \cite{Takai2014,Yamazato2014} have shown that OCC provides V2V communication with the data rate of 10 Mbps. In \cite{Yamazato201723}, the author provides an overview of OCC-based ITS. Table \ref{Tableitocc} summarizes the literature for various OCC-based ITS.

\section{Research Challenges and Future Directions}

Despite various attractive characteristics of OCC, there are several challenging factors to implement practical OCC systems. In this section, we summarize several research challenges and provide future research directions for OCC.

\subsection{Unified System Model and General Performance Analysis}
Since the technical specifications of OCC are not completely standardized, there exists a lot of research on proposing different system architectures with different performance analysis. Therefore, one of the major challenges of OCC is its dependency on the structure of the receivers (cameras), diverse modulation and coding schemes, and different synchronization techniques which cannot be comprehensively compared. Moreover, the current performance analysis criteria in the literature are divergent. Each technique shows its performance where there is no optimality criteria or a performance bound defined to which the results can be compared. Therefore, the research community of OCC is required to develop a unified system model and performance bounds to analyze the performance of different specifications of OCC.

%

\subsection{Modeling of Ambient Light}
Ambient light is one of the primary sources which interfere with OCC. The ambient light modifies the luminance of the received pixels which originate errors in the decoding process and therefore results in an information loss. The ambient light also generates the primary source of noise, i.e., flicker noise which dramatically degrades the performance of the OCC systems. Most of the existing work neglect the effect of ambient light, for example, well known LED2C communication systems such as CeilingTalk system \cite{Yang2017c} and CamComSim \cite{Duque2017a} ignore the impact of ambient light.  One solution to separate the important pixels from the noisy pixels is to use high-speed cameras ($1000$ fps - $10^6$ fps) \cite{Yamazato2016}.


\subsection{Defocusing}
Defocusing is one of the significant challenges faced by OCC where the size of the transmitting source and the camera pixels mismatches even if the source is placed at the camera focus. Therefore, such defocusing is inevitable especially with handheld cameras which are not stable and can cause distortion to each pixel. Moreover, defocusing exists in stationary cases as well which distorts the pixels in the form of mixed frames and blur {\color{red} \cite{Mondal2015}.} Although the impact of blur has been well investigated in the area of computer vision yet the subject on the combined effect of defocusing and blur is an open research problem for OCC.

\subsection{Modeling of Signal Dependent Noise}
The signal dependent shot noise at the receiver in OCC is modeled as a Poisson process due to the nature of the received optical signal.  Also, the shot noise is approximated by the Gaussian model when the received photons are massive in number \cite{optbook}. In this case, the optical channel is modeled by additive white Gaussian noise (AWGN) consisting of both signal depended noise and signal independent noise \cite{Hranilovic2006}. The inclusion of signal dependent noise in OCC systems makes it more challenging as compared to the common AWGN model used in conventional wireless communication systems. Therefore, sophisticated signal processing techniques need to be developed for the transmission and reception of optical signals in OCC systems.

\subsection{Synchronization}
As most of the commercially available cameras have low and diverse frame rates and are unstable, therefore, if the frame rates of the transmitter and the receiver mismatches, the problem of synchronization arises \cite{Hu2013}. For example, the frame rate of a typical smart-phone camera is 30 frames per second while the conventional LCD has the frame rate of 30-60 frames per second. Similarly, the frame rates of LED and screen-based transmitters varies which results in losing the frames or mixing the frames at the receiver. Therefore, novel synchronization techniques are necessary for the development of OCC systems.

\subsection{Hybrid Systems}
The existing LED based communication is considered a significant milestone in the OWC society. In case of LED2C communications, the VLC, Li-Fi and OCC use LED as a transmitter, however the receiver is different in case of OCC (camera). The existing VLC and Li-Fi systems provide high data rate and secure communication but suffers from low SNR and short transmission range \cite{Hasan2018, Chowdhury2018b}. However, OCC can provide long-range communication with high SNR but suffers from low data rate. Therefore, a hybrid system which can support VLC, Li-Fi, and OCC can be good solution to address different applications. Moreover, OCC can also be integrated with RF systems to provide various applications. For example, in \cite{Hasan2018b} a hybrid OCC-RF architecture was presented for real-time patient monitoring. However, a mechanism needs to be devised to switch between the various wireless communications technologies.

\subsection{Robust Image Processing Techniques}
In OCC systems, the received data is in the shape of images and therefore it requires reliable and robust image processing techniques to extract the useful data. Recently, neural and deep networks have been used to solve the complex problems of image recognition and classification. For example, a neural network based equalizer is used in \cite{Zabih2012} to reduce the intersymbol interference for VLC systems. Similarly, neural networks have been used in \cite{Aslam2018} to process images for object detection on the internet of multimedia things (IoMT). As major applications of OCC systems such as localization, navigation, and motion capture depend on image processing tools, therefore sophisticated algorithms need to be devised to improve the performance of all such applications.

\subsection{Medium Access Control}
Most of the existing research work on OCC is focusing on physical layer issues such as modulation, coding, and synchronization. However, other layers such as MAC and network layers are still unexplored. The existing OCC systems are based on the MAC layer protocols of VLC; however, unlike VLC, the OCC systems support mainly unidirectional communication which requires a re-examination of the MAC protocols. Novel MAC protocols need to be developed which can address the different challenges of OCC systems such as low frame rate, defocusing, visual MIMO support, and mobility.
\subsection{Blockage}
Blockage of the Line-of-Sight link is another critical issue for OCC systems due to the mobility of users or obstruction of objects. The received optical intensity is dramatically reduced in such cases which consequently degrades the performance of OCC systems. Although the authors in \cite{Shi2018} have recently attempted to solve this problem, however, they have assumed very short distance (60 cm) between the transmitter and the camera. To the best of author's knowledge, the research on the blockage issue for OCC systems does not exist yet and therefore require attention.
The user's random movements can be explicitly tracked by using Kalman filters to predict the blockage. Also, Markov chain can be used to estimate the probability of blockage in LoS conditions.

\section{Conclusions}
In this paper, we presented a broad survey on optical camera based communications, localization, navigation, and motion capture. This survey covers different features of OCC which includes standardization, modulation and coding techniques, localization and navigation by using OCC, and motion capture with OCC. Firstly, the history and standardization of OCC have been presented followed by the literature on channel characterization, modulation techniques, coding schemes, and synchronization mechanisms. Additionally, OCC has been recently used for localization, navigation, and motion capture for various applications and therefore we have covered these aspects. Besides providing the technical details of physical layer characteristics of OCC, we have introduced several challenges and future research directions. In short, this survey helps the novice readers to understand OCC from its inception to the current developments in multiple perspectives. We believe that with the technological advancement and ongoing research, OCC systems have great potential in the near future.

\section*{References}
\bibliography{IEEEabrv,OCC_reference}

\end{document}